\documentclass[conference]{IEEEtran}
\IEEEoverridecommandlockouts
\usepackage{cite}
\usepackage{amsmath,amssymb,amsfonts}
\usepackage{algorithmic}
\usepackage{graphicx}
\usepackage{textcomp}
\usepackage{xcolor}
\usepackage[capitalise]{cleveref}
\usepackage{flushend}
\usepackage{mathtools}
\usepackage{array}
\usepackage{rotating}
\usepackage{tablefootnote}
\usepackage{ctable}
\usepackage{threeparttable}

\def\BibTeX{{\rm B\kern-.05em{\sc i\kern-.025em b}\kern-.08em
    T\kern-.1667em\lower.7ex\hbox{E}\kern-.125emX}}
\begin{document}

\title{FieldHAR: A Fully Integrated End-to-end RTL Framework for Human Activity Recognition with Neural Networks from Heterogeneous Sensors  
}

\author{\IEEEauthorblockN{
        Mengxi Liu\IEEEauthorrefmark{1},
        Bo Zhou\IEEEauthorrefmark{1}\IEEEauthorrefmark{2},
        Zimin Zhao\IEEEauthorrefmark{1}\IEEEauthorrefmark{2},
        Hyeonseok Hong\IEEEauthorrefmark{3},
        \\Hyun Kim\IEEEauthorrefmark{3},
        Sungho Suh\IEEEauthorrefmark{1}\IEEEauthorrefmark{2},
        Vitor Fortes Rey\IEEEauthorrefmark{1}\IEEEauthorrefmark{2} and
        Paul Lukowicz\IEEEauthorrefmark{1}\IEEEauthorrefmark{2}
        }
    \IEEEauthorblockA{\IEEEauthorrefmark{1}German Research Center for Artificial Intelligence (DFKI), Kaiserslautern, Germany}
    \IEEEauthorblockA{\IEEEauthorrefmark{2}Department of Computer Science, RPTU Kaiserslautern-Landau, Kaiserslautern, Germany}
    \IEEEauthorblockA{\IEEEauthorrefmark{3}Department of Electrical and Information Engineering, Seoul National University of Science and Technology, Korea}
}

\maketitle

\begin{abstract}
In this work, we propose an open-source scalable end-to-end RTL framework FieldHAR, for complex human activity recognition (HAR) from heterogeneous sensors using artificial neural networks (ANN) optimized for FPGA or ASIC integration. 
FieldHAR aims to address the lack of apparatus to transform complex HAR methodologies often limited to offline evaluation to efficient run-time edge applications.
The framework uses parallel sensor interfaces and integer-based multi-branch convolutional neural networks (CNNs) to support flexible modality extensions with synchronous sampling at the maximum rate of each sensor.
To validate the framework, we used a sensor-rich kitchen scenario HAR application which was demonstrated in a previous offline study.
Through resource-aware optimizations, with FieldHAR the entire RTL solution was created from data acquisition to ANN inference taking as low as 25\% logic elements and 2\% memory bits of a low-end Cyclone IV FPGA and less than 1\% accuracy loss from the original FP32 precision offline study.
The RTL implementation also shows advantages over MCU-based solutions, including superior data acquisition performance and virtually eliminating ANN inference bottleneck.

\end{abstract}

\begin{IEEEkeywords}
FPGA, Sensor Fusion, Human Activity Recognition, Neural Networks 
\end{IEEEkeywords}

\section{Introduction}

Human activity recognition (HAR) is an application-oriented discipline that focuses on developing systems capable of inferring the semantic context of human activities from information sources such as sensors using machine learning (ML) algorithms \cite{bian2022human, qiu2022multi}.
HAR has become increasingly relevant with the rise of smart devices, services, and systems, as it enables tailored and context-aware services.
Sensor-based HAR often utilizes ML algorithms, such as pattern recognition and artificial neural networks (ANN), to associate sensor signals with physical activities.
As elaborated in \cref{sec:related_HAR}, the complexity of the real world has resulted in the multi-modal, multifaceted, and temporal-sensitive nature of HAR applications. 
Complementary sensing modalities and sensor fusion are commonly used in HAR to account for the unique sensor outputs associated with different physical activities \cite{qiu2022multi, bharti2018human}.
As human activities are composed of complex sequences of motor movements, capturing these temporal dynamics with a stable high sampling rate is fundamental in HAR \cite{ordonez2016deep}.


With the growth of smart wearable and home devices, HAR has gained interest in edge computing systems, where sensor data acquisition (DAQ), processing, and ML prediction are performed on embedded processors. 
However, while many HAR methodologies have shown promise in offline studies involving heterogeneous sensing of high data quality, few have been transitioned to edge devices for run-time inference in the field, and most are restricted to limited sensors, such as inertial measurement units (IMUs).
Current microprocessor (MCU) architectures struggle with maintaining a high sampling rate or data throughput when more sensor instances, modalities, or larger ML models are deployed to the workload of the same processor. 
As most ML algorithms in HAR are temporal sensitive, maintaining stable sampling rates independent of these system expansions is a basic requirement for run-time HAR systems.
As MCUs execute sequential instructions, increasing sensors may also introduce lag between modalities and simultaneous data collection cannot be guaranteed, which may further negatively impact the recognition result and even lead to catastrophic failure. 
Compared to MCUs, field programmable gate arrays (FPGAs) with many advantages including reconfigurability and parallelism, which support hardware-algorithm co-optimization, have become an interesting embedded platform candidate for complex run-time HAR systems \cite{venieris2021reach}. 
For relatively small systems, FPGAs can also contain all data on-chip, eliminating the bottlenecks of moving data between the off-chip memory \cite{banerjee2021memory}.
However, the knowledge barriers between HAR data science and hardware-specific FPGA application development have so far hindered more edge implementations of complex HAR methodologies, even with available high-level synthesis (HLS) tools \cite{fernandes2019accelerating}.

To overcome these limitations, we propose a fully integrated Register Transfer Level (RTL) end-to-end framework, named FieldHAR, that covers the entire HAR pipeline from DAQ by heterogeneous sensors to activity prediction by ANNs. 
With FieldHAR, the embedded system can reach high performance in both DAQ and ANN inference throughput independent from any system extensions.
In summary, we developed FieldHAR with the following contributions:
\begin{enumerate}
    \item An end-to-end framework from sensor inputs to ANN model activation fully integrated into FPGAs. The framework includes a scalable heterogeneous parallel sensor interface that guarantees the sampling rate and RTL implementation of the ANN model.
    \item An ANN model designed for scalable heterogeneous temporal data based on branched convolutional neural networks (CNNs) for sensor fusion and RTL micro-architectures optimized for its inference. 
    \item Validation with a kitchen scenario HAR application which was demonstrated in an offline study \cite{liu2022smart}. Through resource-aware optimizations and performance evaluations, we demonstrate the effectiveness of FieldHAR in transforming complex offline HAR methodologies to run-time edge systems. 
\end{enumerate}

\section{Related Work}
\subsection{Sensor-based HAR Methodologies}
\label{sec:related_HAR}

In recent years, there has been a considerable amount of work on sensor-based HAR. 
The IMU is one of the most commonly used sensors in HAR applications \cite{kim2019imu, jiang2017feasibility, kundu2018hand}.
Ronao and Cho \cite{ronao2016human} proposed using CNNs to leverage the intrinsic properties of human activities and time-series signals from the accelerometer and gyroscope on a smartphone. 
Their approach enables efficient, effective, and data-adaptive recognition of human activities. 
Apart from the IMU sensors, the electric field-based sensor is also explored in the HAR task.  
Bian et al. \cite{bian2019passive} developed a human body capacitive-based sensor with microwatt-level power consumption to recognize and count gym workouts, which achieved an average counting accuracy of 91\%. 
Cheng et al. \cite{cheng2010active} used conductive textile-based electrodes to measure changes in capacitance inside the human body, by which the human activities, such as chewing, swallowing, speaking, sighing (taking a deep breath), as well as different head motions and positions, can be recognized.
The concurrent use of multiple sensing modalities enjoys many advantages over a single modality\cite{qiu2022multi}, like better robustness and more complex information extraction. 
For example, motion-related activity is usually recognized by analyzing IMU time series; the human physiological information like heart rate, respiratory, and emotion can be extracted by bio-signals like ECG and EEG within a time window.
Thus, multi-modalities sensing and sensor fusion in HAR have become a popular research direction.
Zhang et al. \cite{zhang2020necksense} designed a necklace using multiple sensor data from a proximity
sensor, an ambient light sensor, and an IMU sensor to detect chewing activity and eating episodes. 
Bharti et al. \cite{bharti2018human} proposed a multi-modal and multi-positional system called "HuMan" to recognize and classify the 21 complex at-home activities of humans with results up to 95\%. 
The system consists of practical feature set extraction from specifically selected multi-modal sensor suites, a novel two-level structured classification algorithm that improves accuracy by leveraging sensors in multiple body positions, and improved refinement in the classification of complex activities with minimal external infrastructure support. 
Although many proposed HAR methodologies have demonstrated remarkable performance based on the multiple sensing modalities and efficient neural networks, most of them still stay in offline evaluation on general-purpose computing hardware and lack evaluation of real-world real-time inference on edge devices.

\subsection{Field Implementations of HAR Applications}
\label{sec:FieldApplications}

Field implementations of HAR Applications are crucial for a truly pervasive solution bridging the gap between HAR research and real-world adaptation. 
Although supporting such AI applications on mobile and embedded hardware that is ubiquitous across consumer devices poses important challenges \cite{venieris2021reach},
with the help of the growing ANN frameworks for MCU-based hardware platforms like TensorFlow Lite Micro \cite{david2021TensorFlow}, MicroTVM \cite{chen2018tvm}, CMix-NN \cite{capotondi2020cmix}, CMSIS-NN \cite{lai2018cmsis}, and STM X-Cube-AI \cite{falbo2019analyzing}, more and more works for real-time HAR on MCU-based edge devices have been presented \cite{bian2022exploring,coffen2021tinydl,bian2021capacitive}. 
For example, the work \cite{bian2021capacitive} developed a capacitive-sensing wristband that utilizes four single-end electrodes for onboard hand gesture recognition. By deploying a single convolutional hidden layer as the classifier on the Arduino nano sense platform with a 64 MHz CortexM4 MCU integrated with an FPU, 1 MB flash, 256 KB RAM,
this wristband can accurately identify seven hand gestures from a single user with 96.4\% accuracy in real-time. 
However, the MCU hardware resource constraints often limit more sophisticated implementations from many aspects including data throughput, selection of sensor modalities, and ANN complexity, which are all proven important in offline HAR studies as mentioned in \cref{sec:related_HAR}.

Compare to MCUs, the parallel data processing capability, flexible data representation, and reconfigurability of FPGAs have attracted the attention of many researchers as an alternative hardware platform for field implementations of HAR applications.
Generally, FPGAs provide higher energy efficiency than GPUs and higher performance than CPUs \cite{mittal2014survey}. Existing studies mainly focused on deploying the neural networks on FPGA efficiently \cite{loh2020low,de2020low} or designing a hardware architecture with uniform modality sensing input \cite{mazumder2021automatic}, the former usually requires additional data reading devices, the latter lacks flexibility. The work SensorNet \cite{jafari2018sensornet} also proposed a scalable and low-power embedded CNN for multi-channel time series signal classification, time series from multiple channels were converted to a 2D array, and then the 2D deep CNN was applied to extract features and classify the activities, this architecture can only support sensor fusion from data input level which is not optimal for heterogeneous sensors. 
On the other hand, data acquisition from heterogeneous sensors is a complex task crucial for providing high-quality data input for the ANNs,  and thus shall not be overlooked.
Yet most field implementation studies focus on efficient ANN execution with hardware accelerators \cite{mittal2020survey}.

To the best of our knowledge, our FieldHAR framework is the first complete end-to-end architecture that includes from heterogeneous sensor data acquisition to data processing with ANNs designed for heterogeneous sensor fusion on FPGAs for HAR applications.

\section{Framework Structure}
This framework includes not only a sensor driver hardware library to support flexible extension, rapid implementation, and synchronous sampling at the maximum rate of each sensor but also an adaptive integer-based multi-channel branched CNN that supports both data fusion and feature fusion architecture. 
The open-sourced framework is described by SystemVerilog without using any proprietary IP cores. 
Therefore, it supports flexible migration between different FPGAs or ASICs.

\begin{figure}[!t]
\includegraphics[width=1.0\linewidth]{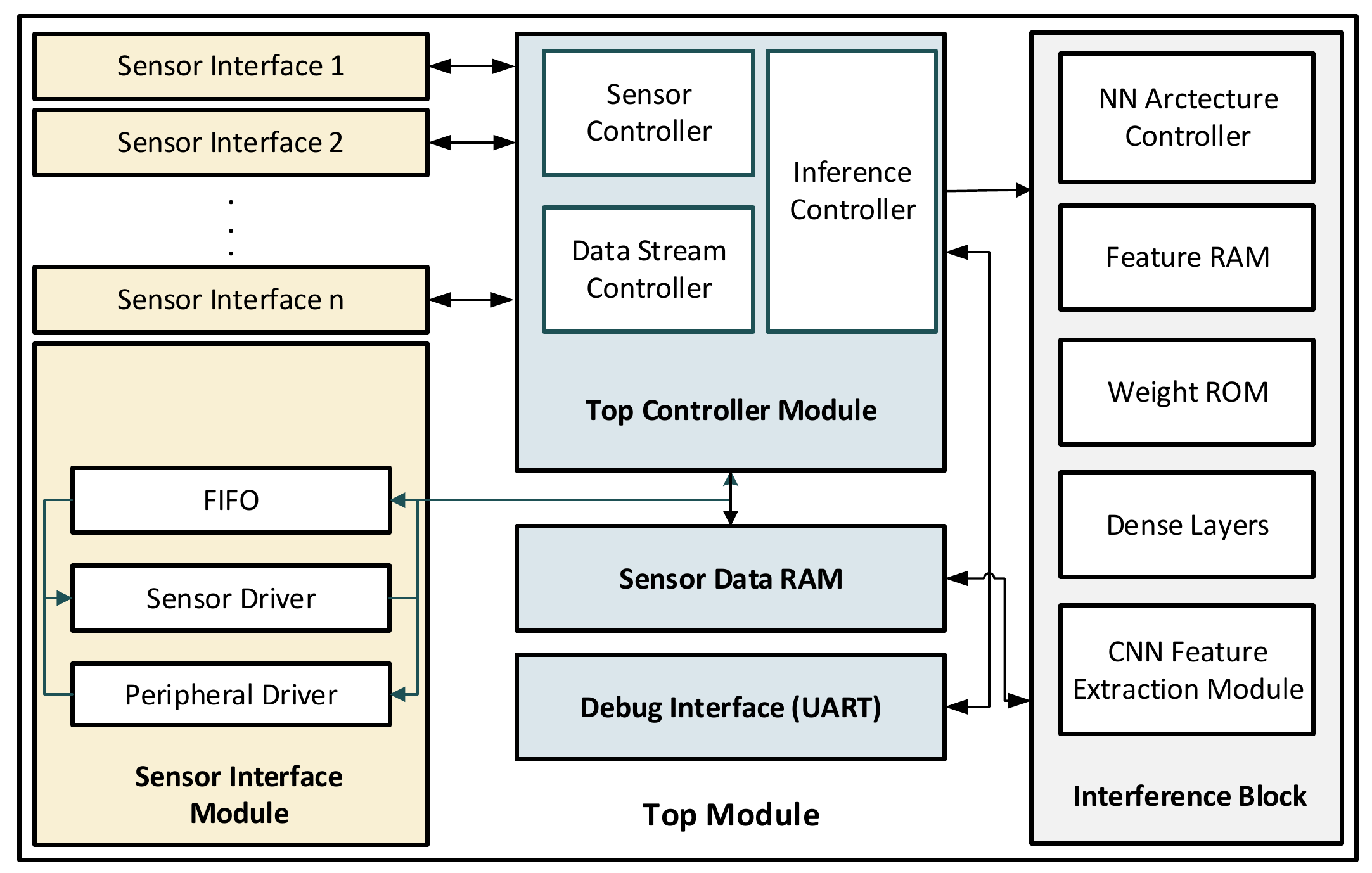}
\caption{The overall structure of FieldHAR}
\label{fig: framework}
\end{figure}

\cref{fig: framework} illustrates the high-level block diagram of our proposed end-to-end RTL framework, which mainly comprises three primary modules: the scalable sensor interface, top controller module, and ANN inference module. Our RTL framework's design is guided by the following objectives:

\begin{itemize}
    \item Fully integrated end-to-end RTL framework: It includes both data acquisition and ML, including automatic feature extraction from heterogeneous sensor data and human activity classification.
    \item Flexibility and scalability: It supports further heterogeneous sensors integrated into the framework easily. 
    \item Resource-efficiency: It supports hardware-algorithms  co-optimization to achieve high resource efficiency. 
\end{itemize}


\subsection{Scalable Sensor Interface}
\begin{figure}[!t]
\includegraphics[width=1.0\linewidth]{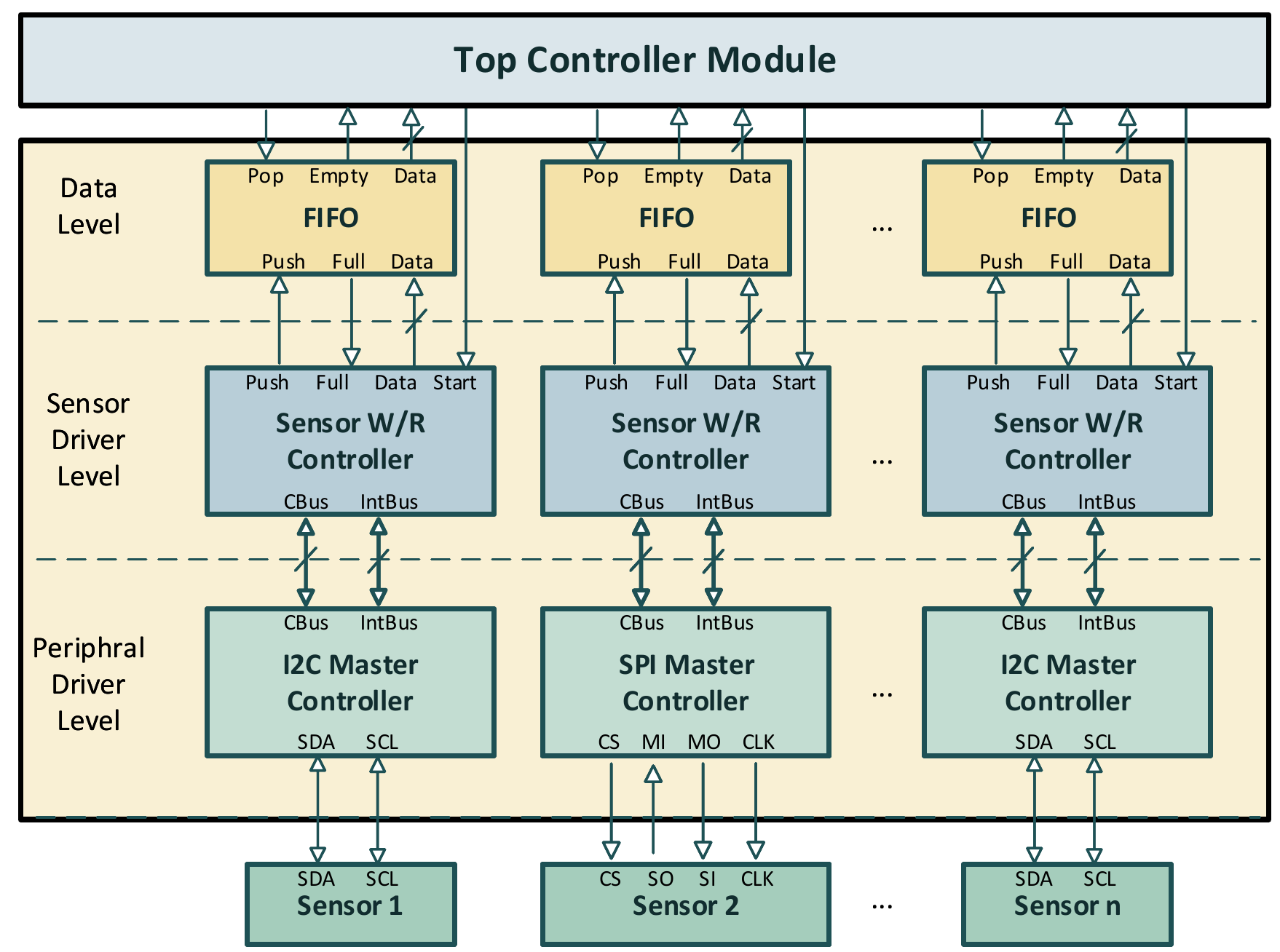}
\caption{Architecture of the parallel sensor interface}
\label{fig:sensorinterface}
\end{figure}

\cref{fig:sensorinterface} shows the architecture of the scalable sensor interface, which is consisted of three levels: peripheral driver level, sensor driver level, and data level. 

At the peripheral driver level, the peripheral driver module directly connects to the sensor and implements the peripheral interface protocol. 
To this end, FieldHAR supports Inter-Integrated Circuit (I2C) and Serial Peripheral Interface (SPI) bus, which are the two primary peripheral protocols used in commercial sensors for HAR.

The sensor driver level performs a function similar to that of the sensor software driver library, which involves two state machines. The first state machine controls data transactions between the I2C/SPI master control modules and the sensors, including single-byte read/write and multiple-byte read/write operations. 
The second state machine completes register operations of sensors, such as control register configuration and sensor status/data registers read. 
As different sensors have distinct register address maps and operation flows, users need to reorder the state transitions and redefine the registered address in the package file when integrating a new sensor into the framework. 
The retrieved sensor data is pushed into the data-level FIFO, and a start signal from the top controller module synchronizes data reads across multiple sensors.
The depth corresponds to the time steps.

\subsection{Top Controller Module}

FieldHAR's workflow is managed by the top controller module with three components:
\begin{itemize}
    \item The sensor controller ensures simultaneous operations among different sensor interfaces.
    \item The data stream controller combines the heterogeneous sensor FIFOs with different sampling rates to a single sensor data RAM.
    \item The interface controller handles ANN activation upon the sensor data RAM ready signal from the data stream controller, and interfaces with external devices via a UART interface, including receiving start/stop commands, sending out inference results or sensor data.
\end{itemize}

In HAR, sliding window is the common approach as there are typically no clear signs of the start and stop of activity instances.
This is implemented with the sensor data RAM so that the window size and step are independent from the individual sensor FIFOs.

\subsection{Neural Networks Inference Module}
\label{sec:ANN_module}
\begin{figure}[!t]
\includegraphics[width=1.0\linewidth ]{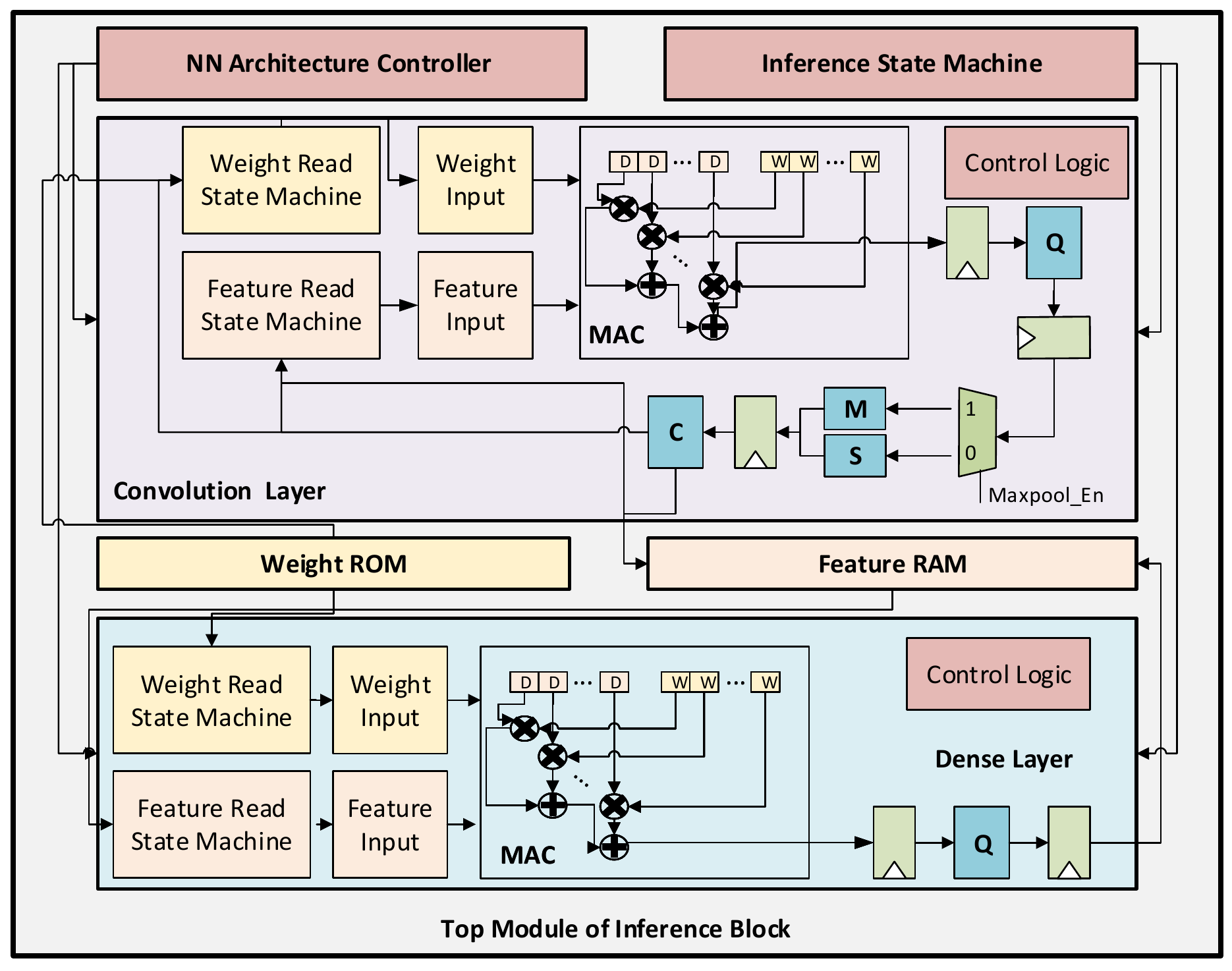}
\caption{Block Diagram of the ANN Inference Module (\textbf{MAC}: multiply-accumulate unit; \textbf{C}: Counter; \textbf{Q}: Quantization; \textbf{S}: Data Shift; \textbf{M}: Global or Kernel Max-pooling)}
\vspace{-10pt}
\label{fig: neural network inference module}
\end{figure}

The ANN inference module is specially designed for a quantized branched CNN feature fusion model native supporting heterogeneous sensors as later discussed in \cref{sec:CNN_structure}.
As shown in \cref{fig: neural network inference module} it comprises a convolution layer module for feature extraction, a dense layer module for classification, and ANN architecture controller.
The ANN model is effectively stored in the weight ROM, and the feature RAM facilitates run-time calculation.

Both the convolution and dense layers consist of a \textit{Weight Read State Machine} and a \textit{Feature Read State Machine} to prime the multiply–accumulate unit (\textit{MAC}) for matrix multiplication.
A quantization (Q) module handles the output requantization required in quantized on-device inference \cite{nagel2021white}.
The non-linear activation (ReLU in this case) is folded inside the Q module.

The convolution layer has an additional shift \textit{S} operation and counter \textit{C} module to facilitate the stepped operation of kernel convolution.
Max pooling (M) of the same kernel size, or global max pooling, is also folded inside the convolution layer by comparators to better utilize the stepped operation, selectable by a multiplexer. 
The convolution kernel size and output channels are implemented in parallel, so the convolution operation scales linearly with input channels.
While the input channels can also be paralleled at the cost of channel-times the resource, our evaluation results in \cref{sec:results} show that the current convolution layer implementation is already providing negligible inference time in HAR applications.
Thus we decide to trade input channel parallelism with hardware resources for more ANN model complexity.

To achieve efficient on-chip memory utilization, resource-aware ANN optimization is applied to reduce the required memory size of the neural networks. 
Firstly, as dense layers take the majority of trainable parameters, using only two dense layer with a small input size after several pooling operations reduces the model size. 
Secondly, quantization \cite{nagel2021white} is applied to reduce the parameter precision and thus data buffer size with negligible performance loss. 
Thirdly, inspired by the work in \cite{mitschke2019fixed}, the bias in the ANN is removed through tensor normalization, which further reduces the trainable parameters. 
These techniques collectively reduce the memory requirements, leading to lower energy consumption and latency by avoiding off-chip memory access during model inference \cite{banerjee2021memory}.

\section{HAR Application-Specific Evaluation}

\subsection{Kitchen Activity Recognition Example}
Monitoring human activity in the kitchen can provide valuable information for improving people's health and well-being. 
By tracking activities such as meal preparation, cooking, and eating, a system can provide personalized advice and guidance to promote healthy eating habits. 
Additionally, monitoring activity in the kitchen can also provide useful information for elderly care, as it allows caregivers to monitor eating patterns and ensure that individuals are receiving adequate nutrition. 
Overall, the kitchen is a critical research area for human activity monitoring and has the potential to improve health outcomes and quality of life.
Thus, a kitchen HAR dataset with multiple sensors acquired from \cite{liu2022smart} was selected as the ANN training dataset to evaluate the proposed framework.

The kitchen HAR dataset is recorded by a DAQ module with six sensors (listed in \cref{tab:sensor_list}) driven by 2 MCUs. 
It contains ten types of kitchen-related activities shown in \cref{tab:activity} performed by ten subjects wearing the DAQ on the chest.
In total, there are 791 channels of sensor data with different sampling rates. 
After synchronization and interpolation, the equivalent sampling rate is 6 Hz (downsampled from 12Hz).

\begin{table}[!t]
\renewcommand{\arraystretch}{1.0}
\centering
\footnotesize
\caption{Sensor List}
\label{tab:sensor_list}
\resizebox{0.98\columnwidth}{!}{
\begin{threeparttable}
\begin{tabular}{c c c c}
\hline
Sensor & Function & Data &  Native Sampling  \\
model & & Channels & Rate (Hz)\\
\hline
AS7431& Optical Spectrum & 10 & 20 Hz $^1$ \\
CCS811& Gas sensor  & 2 & 4 Hz\\
MLX90640 &  Thermal IR (array) & 768 & 32 Hz \\

LPS22HB &  Air pressure sensor & 1 & 75 Hz \\
LSM9DS1& IMU & 9 &119 Hz $^2$ / 20 Hz $^3$\\ 
VL53L0X& ToF ranging sensor & 1 &50 Hz\\
\hline
\end{tabular}
\begin{tablenotes}
\item[1] Recommended Speed.
\item[2] Fastest low power mode
\item[3] Sampling rate of the magnetometer.
\end{tablenotes}
\end{threeparttable}}
\end{table}

\begin{table}[!t]
\renewcommand{\arraystretch}{1.0}
\centering
\footnotesize
\caption{Kitchen Activities in the collected dataset}
\label{tab:activity}
\resizebox{\columnwidth}{!}{
    \begin{threeparttable}
    \begin{tabular}{c l c l }
    \hline
    Activity ID & Activity & Activity ID & Activity\\
    \hline
    1& sitting down &6& opening door  \\
    2& standing up &7& boiling water\\
    3& walking &8 & washing hand\\
    4& opening microwave oven& 9 & cutting food \\
    5& opening freezer&10 $^1$ & drinking beverage\\
    \hline
\end{tabular}
\begin{tablenotes}
\item[1] The five different beverage intake activities are grouped into one class 10
\end{tablenotes}
\end{threeparttable}
}

\end{table}

\subsection{ANN and Sensor Fusion for HAR Task}
\label{sec:CNN_structure}
\begin{figure}[!t]
\includegraphics[width=1.0\linewidth]{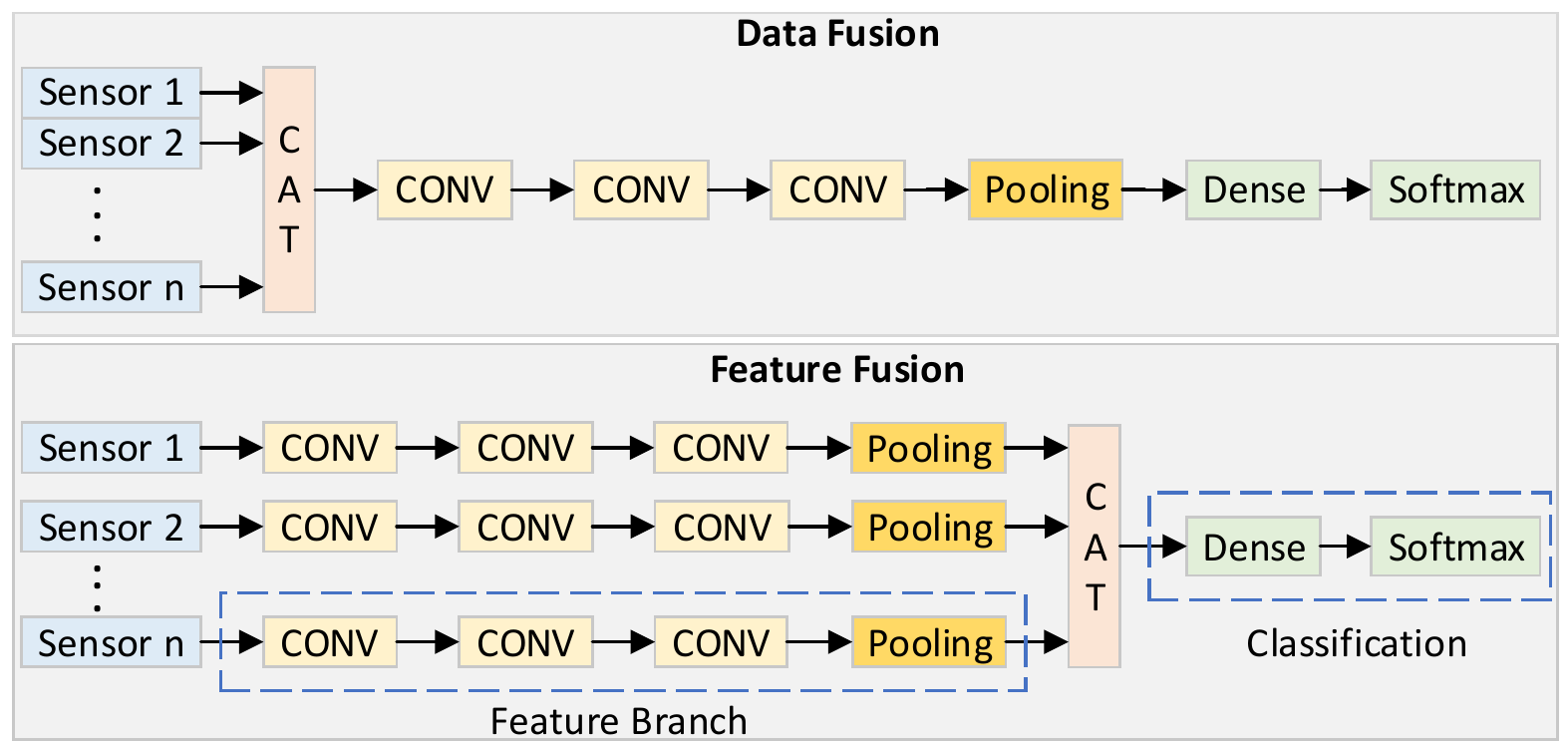}
\caption{The neural architecture of the HAR task with multiple sensor inputs (Global Maxpooling was used).}
\label{fig:NN_architecture}
\vspace{-10pt}
\end{figure}

To classify kitchen activities using data from multiple sensors, two sensor fusion methods were employed in the design of neural networks: data fusion and feature fusion architectures, as depicted in \cref{fig:NN_architecture}. The data fusion architecture is similar to that of SensorNet \cite{jafari2018sensornet}, where time series data from various sensors are concatenated into a two-dimensional matrix (\emph{i.e. }$(W \times C)$, where $W$ and $C$ denote the size of the sliding window and the number of sensor channels, respectively) that inputs to a single neural network branch directly, allowing for simultaneous capture of correlations between various modalities.
When the connected sensors are heterogeneous, for example if one sensor has only one channel while another has several hundred channels, the resulting neural network model may be dominated by the sensor with more channels, leading it to ignore the impact of the sensors with fewer channels or not learn from them at all.
The feature fusion method, on the other hand, uses separate branches of convolution layers extracting features from each sensor.
Thus the imbalanced influence of sensors can be mediated by ensuring similar number of output features per modality. 
The extracted features from each sensor are then concatenated before being fed into the dense layers for classification. 

Feature fusion has shown better accuracy in the literature \cite{munzner2017cnn}, which is also reflected in our evaluation.
An offline experiment with the training data was conducted where two models based on the data fusion and feature fusion methods were built, the result of which are presented in \cref{tab:DF_Comparison}. 
Both models extracted the same number of features, kernel size, and dense layer. 
Thus, the feature fusion architecture was selected for this kitchen activity recognition task, as it offers a higher recognition accuracy with much fewer trainable parameters.

\begin{table}[!t]
\renewcommand{\arraystretch}{1.0}
\centering
\footnotesize
\caption{Performance Comparison of the Kitchen Activity Recognition Between Data Fusion and Feature Fusion Architectures}
\label{tab:DF_Comparison}
\resizebox{\columnwidth}{!}{
    \begin{tabular}{c c c c}
    \hline
    
    \hline
    Sensor Fusion Methods & Features & Trainable Parameter & Accuracy (\%)\\
    \hline
    Data Fusion &28& 71756& 85.43 \\
    Feature Fusion &28&2900& 89.13\\
    \hline

    \hline
\end{tabular}}
\end{table}

In the feature fusion model, data from different sensors were handled by independent feature extraction layers, as shown in the bottom half of \cref{fig:NN_architecture}. 
Each feature branch has three convolution layers with the same filter channels and kernel size, followed by a global max-pooling layer to reduce the temporal dimension to 1.
Then the features are concatenated and fed to two dense layers for classification. 
The softmax activation function of the last dense layer was replaced by a function that outputs the index of the largest output value when deploying this model on the FPGA, which can avoid implementing a division operation on the hardware. The ReLU activation function is used for the rest of the layers.
The filter channels and kernel sizes are hyperparameters that can be adjusted to balance between recognition accuracy and model size. 
For each sensor, independent normalization was applied to rescale the data input range between -1 to 1, which is also prepared for the later quantization step. 
The ANN model was built under TensorFlow 2.10.0 framework, and the model training process was performed on a laptop with the GeForce RTX 3080 Ti GPU. The sparse categorical cross entropy was used as the loss function.

\subsection{Resource-aware Optimizations}
To facilitate efficient ANN deployment onto the FPGA, optimization techniques are employed to reduce the memory and operation footprint of the ANN inference module, including removing less relevant modalities and ANN quantization.

\subsubsection{Modality Selection}
In HAR applications with heterogeneous sensor, it is important to select the modalities that contribute most for the task, as redundant or irrelevant sensors result in unnecessary computational overhead and larger model size. 
To accomplish this, we proposed a method to search for important sensors. In the feature fusion model, there are $n$ parallel feature branches for $n$ sensors, as illustrated in \cref{fig:NN_architecture}. 
Each feature branch outputs a $1\times8$ feature tensor, which we denote as $F_i$. We then assign each sensor a trainable weight $\alpha_i$ that reflects its importance to the classification task. 
These weights are multiplied with corresponding features from each sensor's feature branch and accumulated into a single tensor, $F_{mix}$, for the final classification. 
\begin{equation}
\label{eq: importance_factor}
    F_{mix} = \sum_{i=1}^{n}\frac{\exp\left\{\alpha_i\right\}}{\Sigma_{j=1}^{n}\exp\left\{\alpha_{j}\right\}}F_i
\end{equation}


After the training, we can remove the less useful sensors according to $\alpha_i$ and retrain the model with only the useful sensors without $\alpha_i$. 

For the specific heterogeneous dataset, the IMU data were divided into two categories: motion-related data (accelerator and gyroscope) and magnetic data. 
Besides, 2D convolutions were used to extract the feature from the Thermal IR array as it is analogous to a thermal camera. 
1D convolutions were used for the data from the remaining sensors.

\begin{table}[!t]
\renewcommand{\arraystretch}{1.0}
\centering
\footnotesize
\caption{Sensor Importance Factor of The Training Dataset }
\label{tab:importance_factor}
\begin{tabular}{c c c }
\hline

\hline
Sensor &  Channels  & Importance Factor $\alpha$ $\uparrow$\\
\hline
Optical Spectrum & 10 & 0.240\\
Magnetic (IMU) & 3 & 0.231\\
Motion (IMU) & 6 & 0.180\\
ToF Range & 1 & 0.175\\
Thermal IR (array) & 768 &\textbf{0.098}\\
Gas & 2 & \textbf{0.061}\\
Barometric & 1 & \textbf{0.012} \\
\hline

\hline
\end{tabular}
\end{table}

\begin{figure}[!t]
\includegraphics[width=1.0\linewidth ]{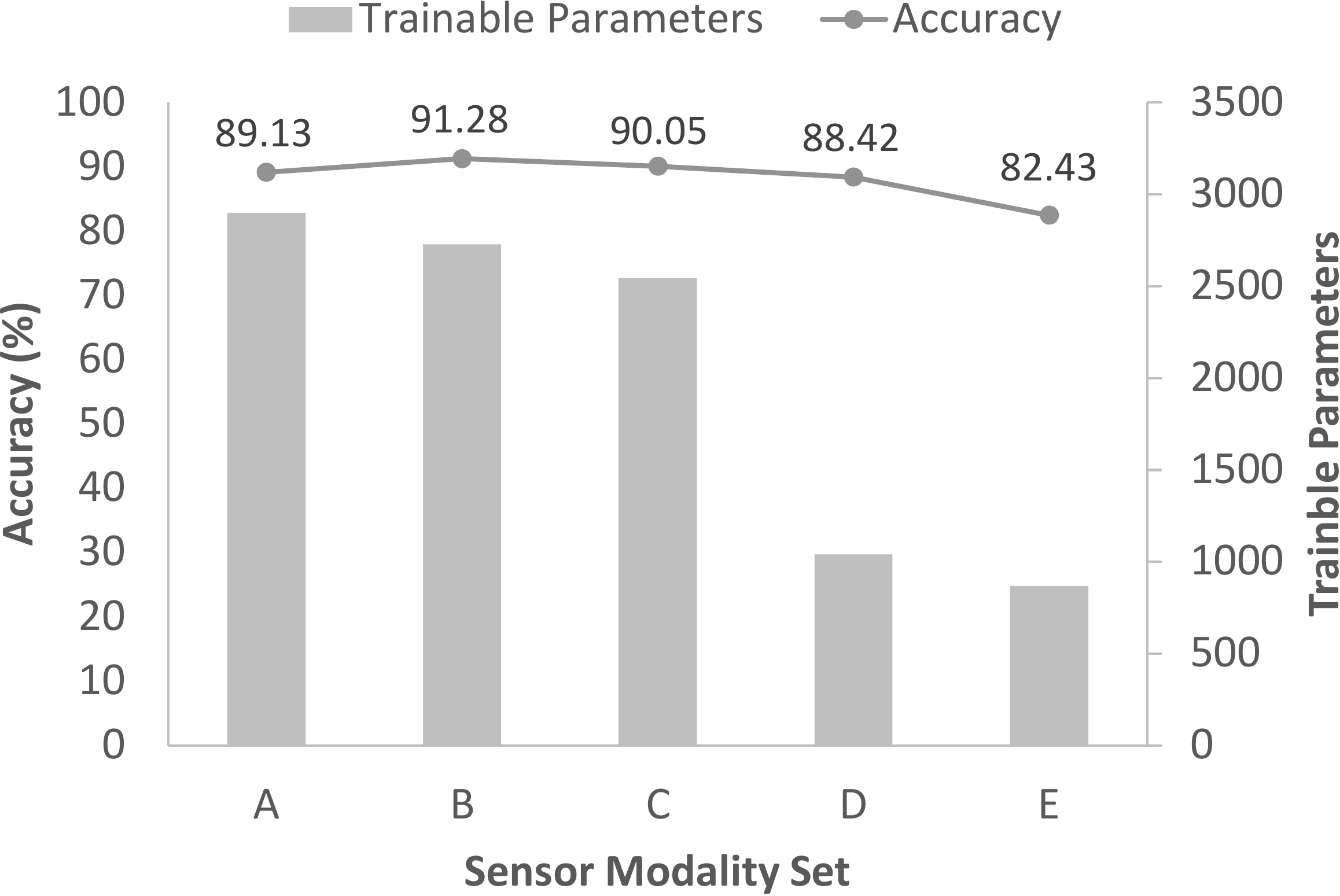}
\caption{Influence of sensor modalities on recognition results and model size. (\textbf{Set A}: includes all seven sensors; \textbf{Set B}: removed Barometric sensor; \textbf{Set C}: removed Barometric and Gas sensors; \textbf{Set D}: removed Barometric, Gas, and Thermal IR sensors; \textbf{Set E}: removed Barometric, Gas, Thermal IR array, and ToF Range sensors)}
\vspace{-10pt}
\label{fig:Sensor_Modality}
\end{figure}

\cref{tab:importance_factor} shows the sensor importance factor of the training dataset. 
To validate the modality selection method, five sensor modality sets were created, where we remove one additional sensor per iteration according from the bottom of the $\alpha_i$  ranking. 
\cref{fig:Sensor_Modality} presents the influence of different sensor modalities on recognition results and model size. 
Despite having less input information, removing the two most insignificant sensors with Set B and C, has even slightly improved the recognition accuracy compared with the full modality Set A.
We find the most cost-effective combination to be Set D with four most significant sensors: the ANN recognition accuracy has a slight decline of around 1\%, while the number of trainable parameters was reduced to almost 1/3 of the full set.

\subsubsection{Post Training Quantization (PTQ)} %
\label{sec:PTQ}
ANN quantization is an effective method for reducing both the model size and computation cost, by which the memory requirement and power consumption of the model during inference can be decreased. 
PTQ specifically does not require retraining the model and thus can be easily adapted.
Reducing the precision from 32-bit to 8-bit could decrease memory resources by a factor of 4 and matrix multiplication cost by a factor of 16 \cite{nagel2021white}.
Given the large number of multiplications and values that need to be stored, such resource savings are crucial when operating CNNs on small or battery-powered edge devices. 
RTL implementations on FPGAs provide even more flexible bit precision options, while MCU-based architectures are usually limited to predefined precision like INT8 or INT16.

PTQ was performed after modality selection, which resulted in a CNN model with four modalities and feature branches.
To find the optimal bit precision, the CNN model was quantized post-training from a FP32 model to n-bit fixed-point integer following the methods in \cite{solovyev2019fixed} with adjustments on tensor normalization to facilitate the branch concatenation of our CNN model. 
The normalization coefficient in the convolution layers was calculated by \cref{eq: rescale_factor} in our work:
\begin{equation}
\label{eq: rescale_factor}
R_l = \max(|W_{l,0}|,|O_{l,0}|,|W_{l,1}|,|O_{l,1}| ... |W_{l,i}|,|O_{l,i}|) 
\end{equation}
where $l$ indicates the CNN layer, $i$ indicates the feature extraction branch, $W$ denotes the weights and $O$ denotes the outputs from the corresponding CNN layer.
As there were three CNN layers from each feature extraction branch, three rescale coefficients $R_l, l=(1,2,3)$ were calculated iteratively. 
This arrangement is to ensure the layer-wise scaling does not change the weight distribution before the concatenation layer.
For the dense layer after concatenating the branched features, normal quantization scaling was performed according to related works \cite{solovyev2019fixed, nagel2021white}. 

Then, the updated weights in fixed-point integer format were calculated according to the symmetric quantization method explained in \cite{nagel2021white} by \cref{eq: weight_update}:
\begin{equation}
\label{eq: weight_update}
W_{int} = \lfloor\frac{W_l}{R_l} \times 2^{n}\rceil
\end{equation}
where $\lfloor\cdot\rceil$ is the operator for rounding to the nearest integer.
$n$ is the quantized bit precision.

\begin{figure}[!t]
\includegraphics[width=1.0\linewidth ]{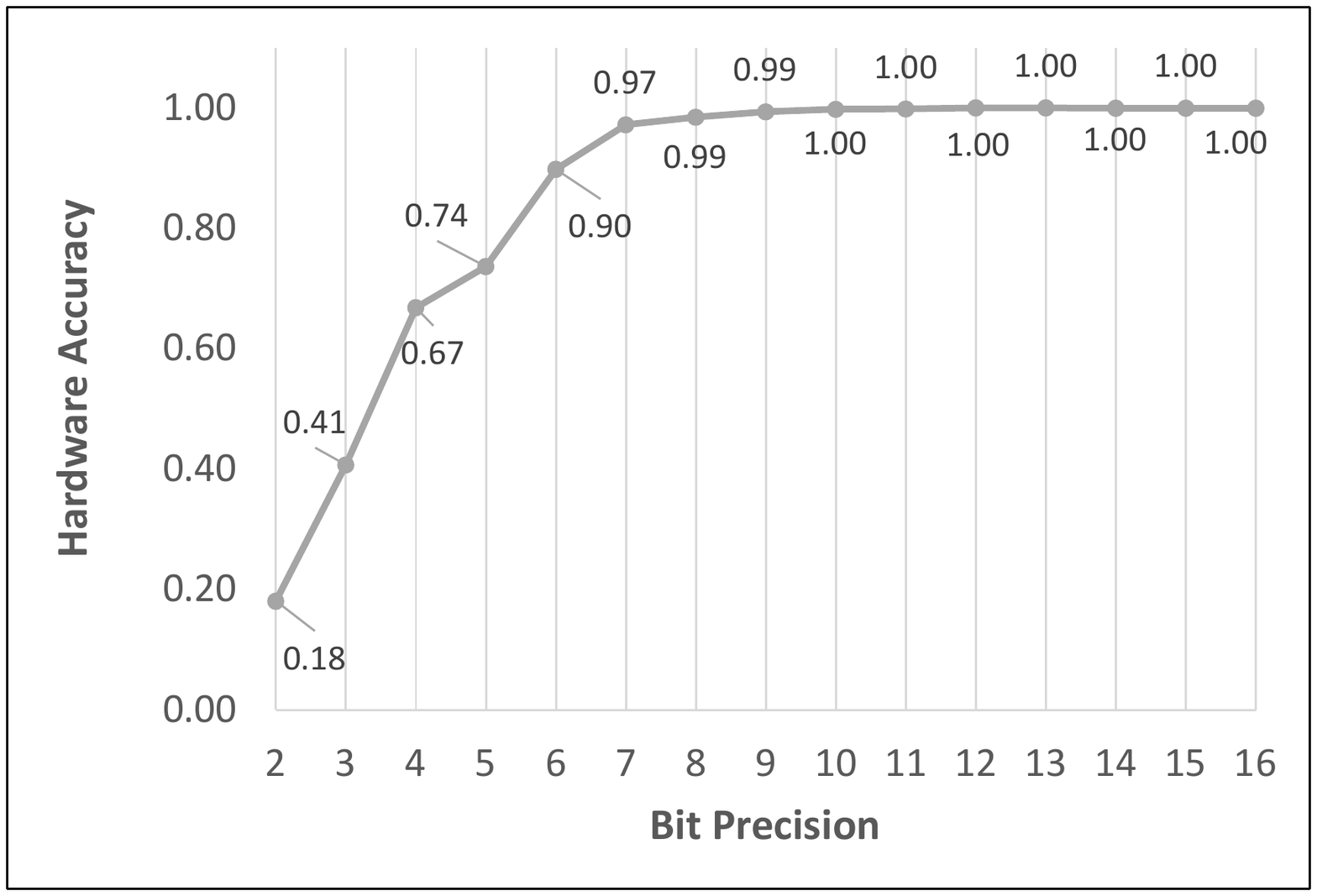}
\caption{The relationship between n-bit quantized accuracy with respect to the FP32 accuracy (without sign bit)}
\vspace{-10pt}
\label{fig:hardware accuracy}
\end{figure}

To evaluate the performance of the model with different quantization bit precision, we use the quantized accuracy / FP32 accuracy as a metric shown in \cref{fig:hardware accuracy}. 
The result indicates that the weight with 10-bit precision can achieve the same accuracy as FP32, and further reducing the bit precision will cause accuracy degradation. 
Although with as low as 7 bits, the accuracy loss of 3\% is still acceptable, the model of 10-bit precision can already be comfortably fit inside our selected FPGA hardware resource as discussed in \cref{sec:results}.  
Thus, the feature fusion neural networks for the kitchen activity recognition task were converted to a 10-bit fixed point format except for one sign bit (signed 11-bit integer). 

\subsection{Parallelism in Model Inference}
The branched feature fusion CNN architecture provides further parallelism potential.
Since each branch is bound to one sensor data source and is independent of each other until the concatenation layer, concurrent computation among these branches can further reduce inference latency.
In addition, as mentioned in \cref{sec:ANN_module}, the convolution layers in this work are designed to leverage the output channel tiling technique as it was identified as the optimal form of parallelism, taking into account both I/O memory bandwidth and computational load, based on the computation-to-communication (CTC) ratio \cite{zhang2015optimizing}.  

\subsection{Hardware Implementation Results and Discussion}
\label{sec:results}
The FieldHAR framework with the kitchen scenario application was implemented on an Intel FPGA Cyclone IV EP4CE22F17C8. 
After optimization, the system has four sensor modalities and the PTQ is set to signed 11-bit integer.
Two types of inference hardware architecture were implemented based on different task schedules: serial and parallel. 
In the serial implementation, feature branches were executed sequentially, while in the parallel implementation, all feature branches were performed in parallel. 
The hardware architecture was described using System Verilog HDL, and the clock frequency was chosen as 100 MHz.

\cref{tab:implementation_result} shows the implementation results for different bit-precision and architectures, indicating a significant impact of the number of precision bits on hardware performance. 
The hardware implementation must have at least an 11-bit precision (including 1 sign bit) to match the FP32 model accuracy as shown in \cref{fig:hardware accuracy}. 
The required logic elements and total memory bits scales almost linearly with the bit precision, showing the flexibility of FPGAs in quantization bit precision as mentioned in \cref{sec:PTQ}. 
However, the multipliers doubled from 9-bit to 11-bit, because the input data width of the hardware-embedded multiplier is 9 bits on the selected FPGA; thus 11-bit operation requires two concatenated multipliers.

\begin{figure}[!t]
\includegraphics[width=1.0\linewidth ]{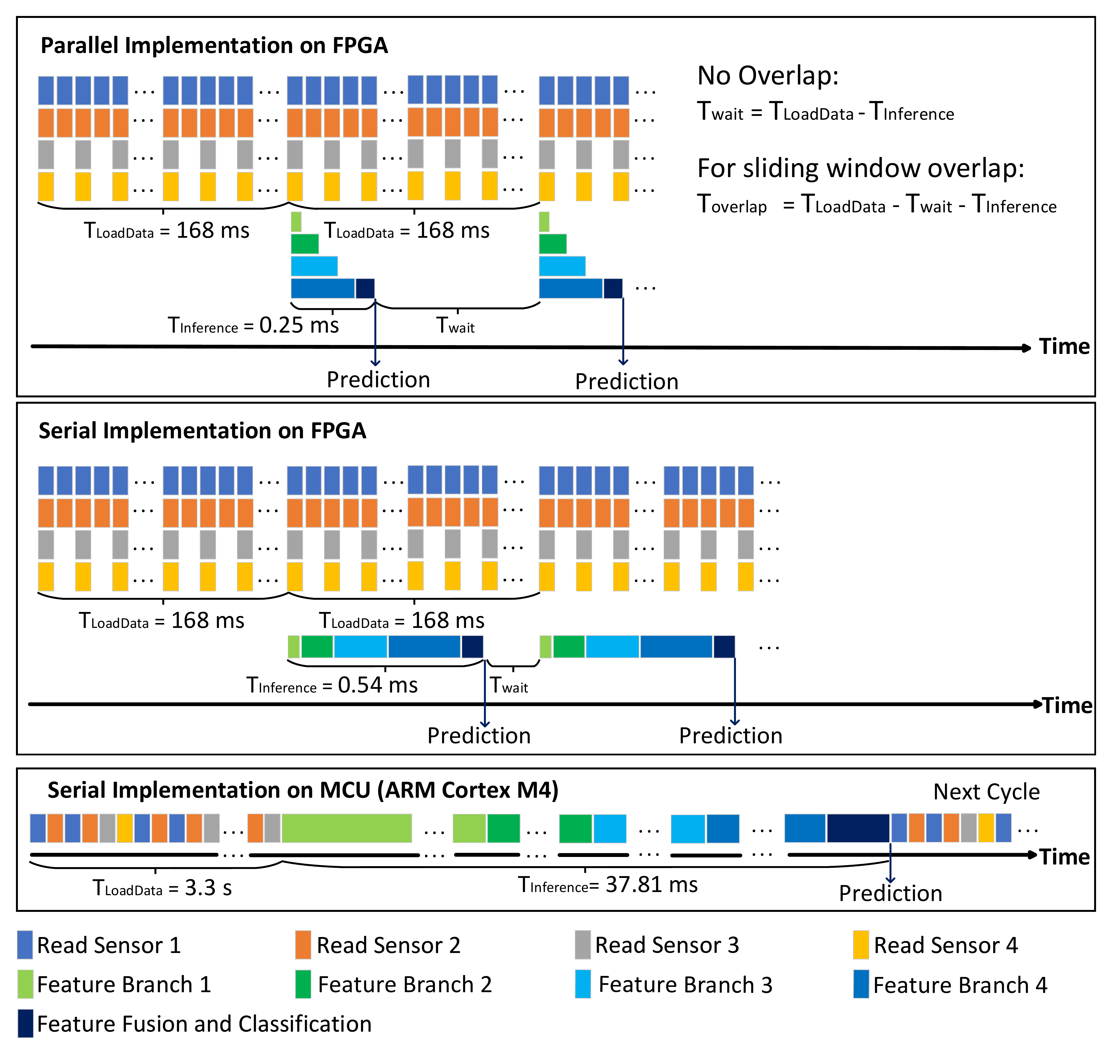}
\caption{Comparison of the HAR Task Schedule between FPGA and MCU, all implementations correspond to 20 samples for the fastest sensor (119Hz possible on the FPGA in this work, and 12Hz possible on the MCU \cite{liu2022smart})}
\label{fig:task_schedule}
\end{figure}

As shown in \cref{fig:task_schedule}, the inference speed has a close relationship with the task schedule strategies, in the serial ANN implementation, the latency of inference was 0.54 ms, while it can be reduced to 0.25 ms by the parallel implementation. 
The fastest throughput of the inference can be up to 4000 labels per second. 
However, the maximum sample rate of most sensors used in HAR is under 1000 Hz, and from the usecase consideration, most recognition for human activities at time window intervals of seconds is already considered fine granularity.
Thus with FieldHAR we can consider the ANN inference is no longer a bottleneck in most HAR applications.
Thus for this specific kitchen scenario application, the 11-bit Serial ANN implementation is already sufficient in terms of latency, while leaving more room for adding more modalities or more complex ANN models in the future.
Serial ANN implementation has less power consumption than parallel implementation because the former design has less hardware resource occupation like logic elements and hardware multipliers. 
In general, the power consumption of the hardware implementation with different configurations (bit precision and parallelism) is under 140 mW, which is slightly more than an ARM Cortex M4 MCU but is suitable for battery-powered edge devices.


\begin{table}[!t]
\renewcommand{\arraystretch}{1.0}
\centering
\footnotesize
\caption{Implementation Result on Intel FPGA Cyclone IV}
\label{tab:implementation_result}
\begin{threeparttable}
\begin{tabular}{c |c c | c c }
\hline

\hline

Metrics/Precision &  \multicolumn{2}{c|}{11 bits } & \multicolumn{2}{c}{9 bits}\\
\hline

\hline
Wn/FP32 Accuracy &  \multicolumn{2}{c|}{100\%} & \multicolumn{2}{c}{99\%}\\
Architecture &  Serial &Parallel& Serial &Parallel\\

\hline
\multicolumn{5}{l}{Inference Block} \\

\hline
Logic Element               & 4473    &13063      &3743      & 10916 \\
(in percentage) & 20\% & 59\% & 17\% & 49\% \\
Total memory bits           & 11440    &35024       &9306     & 28656\\
(in percentage) & 2\% & 6\% & 2\% & 5\% \\

\hline
\multicolumn{5}{l}{Entire System} \\

\hline

Logic Elements               & 6239      &18948       &5501     & 16207 \\

(in percentage) & 28\%           & 85\%       &25\%      & 73\% \\

Total Memory Bits           & 15840     &35024        &12960       & 28659\\
(in percentage)                 & 3\%       & 6\%       &2\%      & 5\% \\
Hardware Multiplier                  & 40            &106         &20        &  53       \\
Clock (MHz)                 & 100           &100         &100       & 100       \\
Latency$^1$  (ms)          & 0.54          &0.25        &0.54      & 0.25      \\
Throughput$^2$  (labels/s)                 & 1851          &4000        &1851      & 4000      \\      
Total Power$^3$ (mW)              & 107.24        &132.67      &106.50    & 124.08    \\
\hline

\hline
\end{tabular}
\begin{tablenotes}
\item[1] Latency of inference block.
\item[2] Throughput of inference block.
\item[3] reported by the Quartus Power Analyzer
\end{tablenotes}
\end{threeparttable}
\end{table}

\subsection{Further discussion and limitations}
From \cref{fig:task_schedule} we can see that the FPGA implementations of FieldHAR can guarantee existing DAQ operations if new tasks, either more sensors or ANN operations, are added, while MCU-based solution struggles in this respect as different tasks need to be scheduled with limited cores.
Even if the tasks can be pipelined with more cores, the FPGA implementation also provides synchrony across modalities.
The training dataset from \cite{liu2022smart} was limited by the MCU during data collection and thus is restricted to 12Hz taking 3.3s for a complete ANN input frame, while the FPGA implementation takes significantly less time (168ms) to collect the input frame.
Even the slower serial ANN is no longer the bottleneck, with 0.54ms latency, 20\% LE and 2\% memory bits.
Thus there is sufficient room for evaluating more complex ANN models with larger input frame with finer time granularity.
Compared with related works with FPGA implementations like \cite{jafari2018sensornet, solovyev2019fixed, mazumder2020energy}, FieldHAR is designed for heterogeneous sensor modalities with different sampling rates, from adaptable sensor interface, branched CNN model with feature fusion, to the optimization step of modality selection; whereas existing works are limited to uniform modality, thus not applicable for the growing sensor fusion based HAR methodologies \cite{qiu2022multi}.

However, the proposed version of FieldHAR to this end has several limitations.
The ANN inference module is limited to convolution, max pooling, concatenation, and dense layers.
Although multi-channel temporal convolution has proven effective in many HAR applications  \cite{qiu2022multi}, there are also other ANN architectures, such as recurrent networks.
The MCU-based platforms mentioned in \cref{sec:FieldApplications} typically support broader selections of layers. 
However, they usually require specific MCU types while FPGA in this regard is more generic. 
While PTQ has already significantly reduced the hardware resource footprint of the ANN model, there are other methods such as quantization-aware training (QAT) that can improve prediction accuracy with lower bits at the cost of additional training for every bit precision.

\section{Conclusion}


In conclusion, FieldHAR presents an end-to-end RTL framework for multi-modal HAR applications, integrating sensor DAQ and ANN model prediction into FPGAs. 
Both the DAQ and ANN modules are designed with modality-wise parallelism through concurrent sensor interfaces and branched CNN models.
The proposed framework is evaluated with a sensor-rich kitchen HAR application scenario from a published offline HAR study. 
Through optimization steps of modality selection and PTQ, we derived a system with four sensors and signed 11-bit integer quantization precision with less than 1\% accuracy loss from the full seven-modality FP32 model.

FieldHAR accommodates the transitions of HAR methodologies which are usually limited with offline evaluations on general purpose computers, to online runtime applications on edge devices.
The parallelism of FPGAs are especially beneficial for multi-modal applications in terms of throughput capability and system robustness against increasing modalities.




\bibliographystyle{IEEEtran}
\bibliography{ref}

\end{document}